\documentclass[%
 reprint,
 amsmath,amssymb,
 aps,
 prl,
 longbibliography,
 lengthcheck,%
]{revtex4-1}

\usepackage{graphicx}% Include figure files
\usepackage{dcolumn}% Align table columns on decimal point
\usepackage{bm}% bold math
\usepackage{hyperref}% add hypertext capabilities
\begin{document}

\preprint{APS/123-QED}

\title{Longitudinal waves in electrically polarized quantum Fermi gas: quantum hydrodynamics approximation}% Force line breaks with \\

\author{P. A. Andreev}
\email{andreevpa@physics.msu.ru}
 \affiliation{Department of General Physics, Physics Faculty, Moscow State
University, Moscow, Russian Federation.}%Lines break automatically or can be forced with \\

\date{\today}% It is always \today, today,

\begin{abstract}
The method of many-particle quantum hydrodynamics has been recently
developed, particularly this method has been used for an
electrically polarized Bose-Einstein condensate. In this paper, we
present the development of this method for an electrically polarized
three dimensional Fermi gas. We derive corresponding dynamical
equations: equation polarization and equation of polarization
current evolution as well as the Euler and continuity equations. We
study dispersion dependencies of collective excitations in a
polarized Fermi gas and consider interference of an equilibrium
polarization on dispersion properties.
\end{abstract}

%\keywords{}%Use showkeys class option if keyword
\maketitle

%\tableofcontents

\section{\label{sec:level1} I. Introduction}

Electrically polarized ultracold Fermi gases have been in the
center of attention ~\cite{Baranov arxiv 07 review}  along with
the electrically polarized Bose-Einstein condensate ~\cite{Baranov
arxiv 07 review}-~\cite{Ni PCCP 09}. It is well-known that
generalization of the Gross-Pitaevskii equation has been used for
description of the electrically polarized Bose-Einstein condensate
what was suggested in Ref.s ~\cite{Yi PRA 00}-~\cite{Santos PRL
00}. An analogous non-linear Schrodinger equation can be used for
study of electrically polarized Fermi gases. However, using of the
kinetic equation is more prevailing (see for example ~\cite{Sogo
NJP 09}, ~\cite{Zhang PRA 09}, ~\cite{Zhang PRA 11}). In simple
cases the set of quantum hydrodynamics (QHD) equations is
equivalent to corresponding non-linear Schrodinger equation
~\cite{Andreev PRA08}, but it we are interested in studying of
evolution of electrical dipole moment direction we need more
general equation set. Such equations can be derived by means of
the QHD method, which allows to get equation of polarization
evolution, as it was demonstrated for electrically polarized
Bose-Einstein condensate ~\cite{Andreev arxiv Pol}-~\cite{Andreev
arxiv 12 transv dip BEC}. This paper is dedicated to both the
development of the method of many-particle QHD for electrically
polarized ultracold Fermi gas and studying of dispersion of bulk
collective excitations inwhere.

To be certain we will consider fermions with the spin 1/2. We
interesting in dynamic of electric polarization, and it's
influence on static and dynamic properties of the ultracold
fermions. We do not consider the dynamic of spin of atoms, for
simplicity. However, at consideration of nonlinear properties of
the electrically polarized  ultracold fermions the role of a spin
could be important. The nonlinear Schrodinger equation is usually
used for quantum gases studying, so it is important to discuss
limits of validity of such approximation. In the case of the
electrically polarized quantum gases, in the corresponding
non-linear Schrodinger equation a new term is added. However, this
approximation includes interaction among parallel dipoles and
influence of this interaction on particles translational motion
(see Fig.1), and does not account spatial and temporal evolution
of dipoles direction pictured on Fig.2. In Ref. ~\cite{Andreev
arxiv Pol} authors developed method accounted spatially
inhomogeneous distribution of dipoles directions and it's temporal
evolution. This method includes the potential of interaction
unparallel dipoles, and it also includes additional equations
determining evolution of dipoles moment of a volume unit. It was
based on the method of many-particle QHD. In our previous papers ~\cite{Andreev arxiv Pol}-
~\cite{Andreev PRB11}  we
developed the method of the many-particle QHD for the electrically
polarized BEC. This approximation allows to describe as the
spatial and the temporal evolution of the electric dipole moments
including evolution of it's directions. Evolution of dipoles
direction gives influence on translational motion of particles due
to tensor nature of the dipole-dipole interaction. Equations of
the QHD are derived directly from the many-particle Schrodinger
equation. QHD description of unpolarized ultracold fermions was
presented in Ref.s ~\cite{Andreev PRA08} and ~\cite{Zezyulin arxiv
12}.

\begin{figure}
\includegraphics[width=8cm,angle=0]{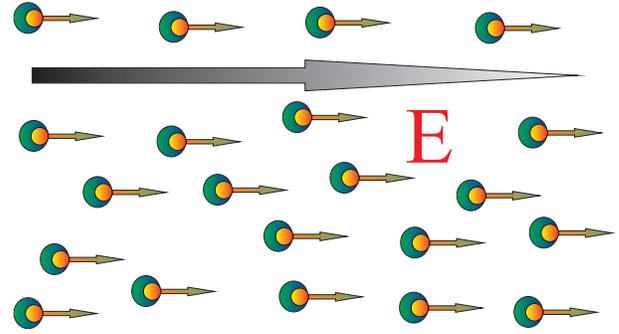}
\caption{\label{dip Fermi parallel} (Color online) The figure
describes the picture of ultracold Fermi particles having electric
dipole moment and obeying to the generalization of the non-linear
Schrodinger equation (the analog of the Gross-Pitaevskii equation
which used for polarized Bose-Einstein condensate description).
This figure  presents system of particles with parallel electric
dipole moment. Arrow by the circle imitates particle dipole and
it's direction.}
\end{figure}

\begin{figure}
\includegraphics[width=8cm,angle=0]{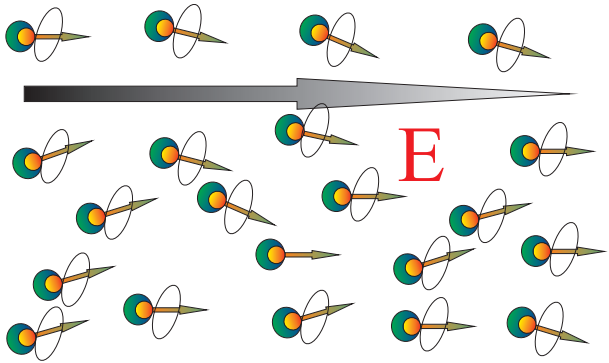}
\caption{\label{dip Fermi rotation} (Color online) The figure
presents the picture of particles having electric dipole moment
where direction of electric dipole moments evaluate in time
(rotating or oscillating) and have different directions in
different points of space (it might be caused by difference of
phase at rotation). Arrow by the circle imitates particle dipole
and it's direction. Round lines around arrow minds rotation of
dipoles.}
\end{figure}

At studying  of ultracold electrically polarized Fermi gas, many
researchers actually prefer using the kinetic equation ~\cite{Sogo
NJP 09}, ~\cite{Zhang PRA 09}, ~\cite{Zhang PRA 11}, which
describes evolution of the distribution function
$f(\textbf{r},\textbf{p},t)$ defined in the six dimensional space
of coordinate $r$ and momentum $p$. The kinetic equation allows us
to get a set of hydrodynamics equations for evolution of the
particle concentration $n(\textbf{r},t)=\int
f(\textbf{r},\textbf{p},t) d\textbf{p}$, momentum density
$\textbf{j}(\textbf{r},t)=\int \textbf{p}
f(\textbf{r},\textbf{p},t) d\textbf{p}$, energy and higher moments
of the distribution function. However the usually used kinetic
equation corresponds to the hydrodynamic equations, which do not
include the equations of polarization evolution, and, so it
describe polarization evolution caused by changing of particles
concentration only. We believe that such approximation might be
enough for getting description of collective excitation of the sample shape, but it is not enough for
bulk waves description in the electrically polarized Fermi gas.

In Ref.s ~\cite{Sogo NJP 09}, ~\cite{Zhang PRA 09} the kinetic equation was derived via the
Wigner function with parallel dipoles without dynamics of electric
dipoles direction. In Ref. ~\cite{Zhang PRA 11} this equation has
been used for dynamics of the dipolar Fermi gas at finite
temperature. The functional renormalization group technique used
in Ref. ~\cite{Bhongale PRL 12} to get  the zerotemperature phase
diagram of dipolar fermions on a two-dimensional square lattice at
half filling for the system of dipoles pointing in the same
direction.

A. R. P. Lima and A. Pelster ~\cite{Lima PRA 10} presented the
derivation of the continuity and Euler equations for dipoles
fermions considering the evolution of the one-body density matrix
~\cite{Bruun PRL 99}-~\cite{Ring book 04}. They considered the set of
equations which is equivalents to the corresponding non-linear Schrodinger
equation and describes dynamics of parallel dipoles. In QHD description
these equations are only a part of the set of QHD equations. Here we
present more general set of QHD equations which fully described
dynamics of electrical polarization in polarized Fermi gas. The Habbard's model is also very useful ~\cite{Lu 11 arxiv},
~\cite{Gadsbolle 11 arxiv}, but we are not going to discuss it here.

%The continuity and the
%Euler equations are the part of the chain in the set of the QHD
%equations. The continuity and the Euler equations correspond to
%the scalar nonlinear Schrodinger equation ~\cite{Yi PRA
%00}-~\cite{Santos PRL 00}, ~\cite{Andreev PRA08}.

The set of the many-particle QHD equations obtained in Ref.s
~\cite{Andreev arxiv Pol}- ~\cite{Andreev RPJ 12} contains four
equation, which are the continuity equation, the Euler equation
(the momentum balance equation), the equation of polarization
evolution and the equation of polarization current evolution.
These equations appear along with the equations of field, which
are $\nabla \textbf{E}=-4\pi\nabla \textbf{P}$ and $\nabla\times
\textbf{E}=0$, where $\textbf{E}$ is the electric field, and
$\textbf{P}$ is the density of electric dipole moment, they are a
pair of the Maxwell's equations. This model give us possibility to
study the evolution of polarization, and it's influence on
particle motion. It was found that in linear approximation where
are two wave solution instead of the Bogoliubov's mode existing in
the unpolarized Bose-Einstein condensate.

In this paper we develop the many-particles QHD for the
electrically polarized  ultracold fermions. We derive
corresponding the continuity equation, the Euler equation, the
equation of polarization evolution and the equation of
polarization current evolution. Expecting to obtain two wave
solutions instead of one existing in the unpolarized ultracold
Fermi gas we consider linear approximation of the QHD equations.

This paper is organized as follows. In Sec. II we present the
derived set of the QHD equations for the electrically polarized
ultracold fermions. In Sec. III we describe the method used for
the solving of the QHD equations and present formula for
dispersion dependence of waves in the system of the electrically
polarized  ultracold fermions. In Sec. IV we present numerical
analysis of the dispersion dependence.  In Sec. V we present the
brief summary of our results.

\section{\label{sec:level1} II. Basic equations}

Here we will briefly present the set of the QHD equations for the
electrically polarized  ultracold Fermi gas. The method of theirs
derivation described in Ref.s ~\cite{Andreev arxiv 12 (2)} and
~\cite{Andreev PRB11}. It was done for the Bose-Einstein
condensate, but the method of derivation is the same, this is why
we do not present details of derivation and present resulting
equations only.

The first equation of the QHD equations system is the continuity
equation
\begin{equation}\label{DF3D cont eq}\partial_{t}n+\partial^{\alpha}(nv^{\alpha})=0,\end{equation}
where $n$ is the particles concentration and $v^{\alpha}$ is the velocity field.

The momentum balance equation for the polarized ultracold fermions has the form
$$mn(\partial_{t}+\textbf{v}\nabla)v^{\alpha}+\frac{\hbar^{2}}{5m}(3\pi^{2})^{\frac{2}{3}}\partial^{\alpha}n^{\frac{5}{3}}$$
$$-\frac{\hbar^{2}}{4m}\partial^{\alpha}\triangle
n+\frac{\hbar^{2}}{4m}\partial^{\beta}\Biggl(\frac{\partial^{\alpha}n\cdot\partial^{\beta}n}{n}\Biggr)
$$
$$=-\Upsilon_{2}(\partial^{\alpha}n)\triangle n-2\Upsilon_{2}(\partial^{\alpha}\partial^{\beta}n)\partial_{\beta}n$$
\begin{equation}\label{DF3D bal imp eq short}+3\Upsilon_{2}n\partial^{\alpha}\triangle
n-4(3\pi^{2})^{\frac{2}{3}}\Upsilon_{2}\partial^{\alpha}n^{\frac{8}{3}}+P^{\beta}\partial^{\alpha}E^{\beta},
\end{equation}
where
\begin{equation}\label{DF3D
Upsilon2}\Upsilon_{2}\equiv\frac{\pi}{30}\int dr
(r)^{5}\frac{\partial U(r)}{\partial r}.\end{equation}
In equation
(\ref{DF3D bal imp eq short})  we defined a parameter
$\Upsilon_{2}$ as (\ref{DF3D Upsilon2}). This definition differs
from the one in Ref.s ~\cite{Andreev PRA08}, ~\cite{Zezyulin arxiv 12}. In left-hand side of
equation we have three terms proportional to $\hbar^{2}$, first of
them is the gradient of the Fermi pressure. Other two are the
quantum Bohm potential, they appear as a result of using of the
quantum kinematics. The first four terms in the right-hand side of
equation (\ref{DF3D bal imp eq short}) describe the short range
interaction in the system of ultracold fermions appearing in the
third order by the interaction radius. They occur because of
taking into account of the SRI potential $U_{ij}$. The interaction
potential $U_{ij}$ defines the macroscopic interaction constant
$\Upsilon_{2}$. The first three terms in the right-hand side of
equation contain high space derivatives of particle concentration
and, thus, have something common with the quantum Bohm potential.
The fourth term arises because of the dependence of the
interaction on the Fermi pressure. Therefore the Fermi pressure
gives contribution in two terms, kinetic and dynamical. The last
term in the equation (\ref{DF3D bal imp eq short}) describes force
field which affects the dipole moments in a unit of volume as the
effect of the external electrical field and the field produced by
other dipoles. The last term is written using the self-consistent
field approximation ~\cite{Andreev PRB11}.

We have also field equations
\begin{equation}\label{DF3D field eq div}\nabla\textbf{E}(\textbf{r},t)=-4\pi \nabla\textbf{P}(\textbf{r},t),\end{equation}
and
\begin{equation}\label{DF3D field eq rot}\nabla\times\textbf{E}(\textbf{r},t)=0.\end{equation}
Equations (\ref{DF3D field eq div}) and (\ref{DF3D field eq rot})
allow us to consider longitudinal waves only, i.e. electric field
of  wave parallel to the direction of propagation.

In the case particles does not contain the dipole moment, the
continuity equation and the momentum balance equation form a
closed system of equations. When the dipole moment is taken into
account in a momentum balance equation, a new physical value
emerges, a polarization vector field $P^{\alpha}(\textbf{r},t)$.
This causes system of equations to become incomplete.

We need next equation for investigation of the dispersion of the collective excitations is the
equation of polarization evolution
\begin{equation}\label{DF3D eq polarization}\partial_{t}P^{\alpha}+\partial^{\beta}R^{\alpha\beta}=0,\end{equation}
$R^{\alpha\beta}(\textbf{r},t)$ is the current of polarization.

The equation (\ref{DF3D eq polarization}) does not contain
information about the effect of the interaction on the
polarization evolution. The evolution equation of
$R^{\alpha\beta}(\textbf{r},t)$ can be constructed by analogy with
the above derived equations.  Using a self-consistent field
approximation of the dipole-dipole interaction we obtain an
equation for the polarization current
$R^{\alpha\beta}(\textbf{r},t)$ evolution
 $$\partial_{t}R^{\alpha\beta}+\partial^{\gamma}\biggl(R^{\alpha\beta}v^{\gamma}+R^{\alpha\gamma}v^{\beta}-P^{\alpha}v^{\beta}v^{\gamma}\biggr)$$
$$+\frac{1}{m}\partial^{\gamma}r^{\alpha\beta\gamma}-\frac{\hbar^{2}}{4m^{2}}\partial_{\beta}\triangle P^{\alpha}$$
\begin{equation}\label{DF3D eq for pol current gen selfconsist
appr}+\frac{\hbar^{2}}{8m^{2}}\partial^{\gamma}\biggl(\frac{\partial_{\beta}P^{\alpha}\partial_{\gamma}n}{n}+\frac{\partial_{\gamma}P^{\alpha}\partial_{\beta}n}{n}\biggr)
=\frac{\sigma}{m}\frac{P^{\alpha}P^{\gamma}}{n}\partial^{\beta}E^{\gamma}.\end{equation}
Here $r^{\alpha\beta\gamma}$ is an analog of the tensor of kinetic
pressure. Assuming to the fact that we consider the system of the
ultracold fermions, and the pressure $p$ described by the Fermi
pressure (the second term in the left-hand side of equation
(\ref{DF3D bal imp eq short})). Therefore we suggest following
equation of state for the $r^{\alpha\beta\gamma}$
\begin{equation}\label{DF3D eq of state} r^{\alpha\beta\gamma}=\kappa\hbar^{2}(3\pi^{2})^{2/3}n^{2/3}P^{\alpha}\delta^{\beta\gamma}/5m,\end{equation}
where $\kappa$ is a numerical constant. Functionally
$r^{\alpha\beta\gamma}$ depends on concentration
$r^{\alpha\beta\gamma}\sim n^{2/3}$ and polarization
$r^{\alpha\beta\gamma}\sim P^{\alpha}$. If we consider a system of
particles with parallel electric dipole moments, which direct
along $z$ axis, we get that $P^{\alpha}=p_{0}^{z}
n\delta^{z\alpha}$, where $p_{0}^{z}$ is the
 constant electric dipole moment of single particle. Thus $P^{\alpha}$
 changes due to change of concentration $n$ only. This approximation
 corresponds to using of the non-linear Schrodinger equation. In this
 case we also get that $R^{\alpha\beta}=p_{0}^{z} n v^{\beta}\delta^{z\alpha}$
 and equation (\ref{DF3D eq for pol current gen selfconsist appr}) reduces to
 the Euler equation (\ref{DF3D bal imp eq short}) where in the last term we
 should put $P^{\alpha}=p_{0}^{z} n\delta^{z\alpha}$. From this comparison we
 conclude that $\kappa$ should be equal to $1$, but we will keep quantity
 $\kappa$ through the paper to trace contribution of the Fermi pressure
 appearing in the equation of the polarization current evolution. The last
term in the formula (\ref{DF3D eq for pol current gen selfconsist
appr}) includes both external electrical field and a
self-consistent field that particle dipoles create. This term
contains a numerical constant $\sigma$.

Equation (\ref{DF3D eq for pol current gen selfconsist appr}) does
not contain short range interaction because  the short range
interaction gives no contribution in equation (\ref{DF3D eq for
pol current gen selfconsist appr}) in the first order by the
interaction radius (analogously to equation (\ref{DF3D bal imp eq short})),
and we do not consider contribution of higher
order. For comparison we can admit that in the Euler equation
(\ref{DF3D bal imp eq short}) we have accounted the short range
interaction up to the third order by interaction radius.

Described equations correspond to the many-particles microscopic
Schrodinger equation, where for dipole-dipole interaction we have used next formula
$$H_{dd}=-\partial^{\alpha}\partial^{\beta}\frac{1}{r}\cdot d_{1}^{\alpha}d_{2}^{\beta}.$$
Which differs from usually used Hamiltonian of dipole-dipole
interaction by the term proportional to the Dirac's delta function.

We present and use the set of the QHD equations in the form it has
been derived. Thus we consider equations of field (\ref{DF3D field
eq div}) and (\ref{DF3D field eq rot}). In Ref. ~\cite{Andreev
arxiv 12 transv dip BEC}, at studying of electrically polarized
Bose-Einstein condensate, the QHD equations was considered along
with the whole set of Maxwell's equation, what allowed consider
transverse waves in polarized quantum gases. In the result it was
shown that electromagnetic waves splits on two branches, and
matter waves also contain transverse component that leads to
anisotropy of the spectrum of collective excitations, whereas
using of equations (\ref{DF3D field eq div}) and (\ref{DF3D field
eq rot}) allows to consider longitudinal waves only what gives no
anisotropy.

\section{\label{sec:level1} III. Elementary excitations in the electrically polarized Fermi gas}

We can analyze the linear dynamics of the collective excitations in the
polarized BEC using the QHD equations (\ref{DF3D cont eq})-
(\ref{DF3D eq of state}). In the beginning we consider
the system is placed in the external electrical field parallel to direction of wave propagation
$\textbf{E}_{0}=E_{0}\textbf{e}_{x}$. Below we will consider influence of the equilibrium external electric field $\textbf{E}_{0z}=E_{0z}\textbf{e}_{z}$, so we will have $\textbf{E}_{0}=E_{0x}\textbf{e}_{x}+E_{0z}\textbf{e}_{z}$. The values of concentration
$n_{0}$ and polarization $\textbf{P}_{0}=\kappa\textbf{E}_{0}$ for
the system in the equilibrium state are constant and uniform and
its velocity field $v^{\alpha}(\textbf{r},t)$ and tensor
$R^{\alpha\beta}(\textbf{r},t)$ values are zero.

We consider the small perturbation of equilibrium state like
$$\begin{array}{ccc}n=n_{0}+\delta n,& v^{\alpha}=0+v^{\alpha},& E^{\alpha}=E_{0}^{\alpha}+\delta E^{\alpha} \end{array}$$
\begin{equation}\label{DF3D equlib state BEC}\begin{array}{ccc}& & P^{\alpha}=P_{0}^{\alpha}+\delta P^{\alpha}, R^{\alpha\beta}=0+\delta R^{\alpha\beta}.\end{array}\end{equation}
Substituting these relations into system of equations (\ref{DF3D
cont eq})- (\ref{DF3D eq of state})
and neglecting nonlinear terms, we obtain the set of
linear homogeneous equations in partial derivatives with constant
coefficients. Passing to the following representation for small
perturbations $\delta f$ in the form of plane wave propageting in direction of $x$ axis
$$\delta f =f(\omega, k) exp(-\imath\omega t+\imath k x) $$
yields the homogeneous system of algebraic equations. The electric
field strength is assumed to have a nonzero value. Expressing all
the quantities entering the system of equations in terms of the
electric field, we come to dispersion equation. Equation (\ref{DF3D field eq div}) gives us the dispersion equation and equation (\ref{DF3D field eq rot}) gives us additional condition on $\delta E_{y}=0$, $\delta E_{z}=0$, so all quantities we express via $\delta E_{x}$. Thus, in presented model we have deal with longitudinal wave as self-consistent electric field in the wave parallel to the direction of wave propagation.  Solving this equation we get despersion dependesies. Here we discuss dispersion equation emerging from evolution of $\delta n$, $\delta v_{x}$, $\delta P_{x}$, $\delta R_{xx}$ and $\delta E_{x}$.

The dispersion characteristic for collective excitations in the electrically polarized
Fermi gas can be expressed in the form of
$$\omega^{2}=\frac{1}{2}\Biggl(\frac{\hbar^{2}k^{4}}{2m^{2}}+\biggl(\frac{1}{3}+\frac{\kappa}{5}\biggr)\frac{\hbar^{2}}{m^{2}}(3\pi^{2})^{2/3}n^{2/3}k^{2}$$
$$+\frac{3\Upsilon_{2}n_{0}k^{4}}{m}+ \frac{4\pi\sigma
P_{0}^{2}k^{2}}{m
n_{0}}+\frac{32}{3}\frac{(3\pi^{2}n_{0})^{2/3}}{m}n_{0}k^{2}\Upsilon_{2}$$
$$\pm\Biggl(\biggl(\biggl(\frac{1}{3}-\frac{\kappa}{5}\biggr)\frac{\hbar^{2}}{m^{2}}(3\pi^{2}n_{0})^{2/3}k^{2}+\frac{3\Upsilon_{2}n_{0}k^{4}}{m}$$
$$-\frac{4\pi\sigma P_{0}^{2}k^{2}}{m
n_{0}}+\frac{32}{3}(3\pi^{2}n_{0})^{2/3}\frac{n_{0}}{m}k^{2}\Upsilon_{2}\biggr)^{2}$$
\begin{equation}\label{DF3D general disp dep}+32\pi\kappa(3\pi^{2})^{2/3}\frac{\hbar^{2}}{15 m^{2}} \frac{k^{4}P_{0}^{2}}{m n_{0}}\Biggr)^{1/2}\Biggr).\end{equation}
The last term under the square root, which is the last term in
formula (\ref{DF3D general disp dep}), is positive. Other term
under the square root is the square of sum of terms, so we have
the sum of two positive quantities under the square root. This
means that $\omega^{2}$ has no imaginary part for considered
system of particles and we can conclude that there are no linear
instabilities in this case.

For numerical analyzes of obtained formula (\ref{DF3D general disp
dep}) we introduce dimensionless parameters $\Omega\equiv m\omega/
(\hbar n_{0}^{2/3}) $, $\gamma=mn_{0}\Upsilon_{2}/\hbar^{2}$, $d=m
P_{0}^{2}/(\hbar^{2}n_{0}^{5/3}) $, and $\xi=k/n_{0}^{1/3}$.

Let's rewrite formula (\ref{DF3D general disp dep}) in
dimensionless variables
$$\Omega^{2}=\xi^{2}\Biggl(\frac{1}{4}\xi^{2}+\frac{3}{2}\gamma\xi^{2}+\frac{1}{2}\biggl(\frac{1}{3}+\frac{\kappa}{5}\biggr)(3\pi^{2})^{2/3}$$
$$+2\pi\sigma d+\frac{16}{3}(3\pi^{2})^{2/3}\gamma  \pm\frac{1}{2}\Biggl(\biggl((3\pi^{2})^{2/3}\biggl(\frac{1}{3}-\frac{\kappa}{5}\biggr)$$
$$-4\pi\sigma d+\frac{32}{3}(3\pi^{2})^{2/3}\gamma+3\gamma\xi^{2}\biggr)^{2}$$
\begin{equation}\label{DF3D general disp dep dimless}+\frac{32\pi\kappa}{15}(3\pi^{2})^{2/3}d\xi^{2}\Biggr)^{1/2} \Biggr).\end{equation}
The first term in this formula appears from the quantum Bohm
potential, the second and nine terms exist due to SRI which we
account up to the third order by interaction radius. The third and
sixth terms are caused by the Fermi pressure which is a part of
equations (\ref{DF3D bal imp eq short}) and (\ref{DF3D eq for pol
current gen selfconsist appr}) via kinetic terms. The fourth and
seventh terms present contribution of equilibrium polarization.
The fifth and eighth terms appear as consequence of contribution
of the Fermi pressure in the SRI. The last term in formula
(\ref{DF3D general disp dep dimless}) appears due to simultaneous
account of the equilibrium polarization and the Fermi pressure.

First of all we should consider the limit of a small polarization
contribution, to compare it with dispersion of unpolarized quantum
Fermi gases. Making series of the square root in formula (\ref{DF3D general disp dep dimless})
on small parameter $d$ up to the linear terms we get
$$\Omega^{2}=\frac{1}{3}(3\pi^{2})^{2/3}\biggl(1+32\gamma\biggr)\xi^{2}+\biggl(\frac{1}{4}+3\gamma$$
\begin{equation}\label{DF3D disp dep dimless small pol Fermi plus}+\frac{8\pi\kappa}{15}\frac{(3\pi^{2})^{2/3}d}{(3\pi^{2})^{2/3}\biggl(\frac{1}{3}-\frac{\kappa}{5}\biggr)+\frac{32}{3}(3\pi^{2})^{2/3}\gamma+3\gamma\xi^{2}}\biggr)\xi^{4},\end{equation}
for the solution with the plus infront of the square root, and
$$\Omega^{2}=\frac{1}{3}(3\pi^{2})^{2/3}\biggl(\frac{3\kappa}{5}+4\pi\sigma d\biggr)\xi^{2}+\biggl(\frac{1}{4}$$
\begin{equation}\label{DF3D disp dep dimless small pol Fermi minus}-\frac{8\pi\kappa}{15}\frac{(3\pi^{2})^{2/3}d}{(3\pi^{2})^{2/3}\biggl(\frac{1}{3}-\frac{\kappa}{5}\biggr)+\frac{32}{3}(3\pi^{2})^{2/3}\gamma+3\gamma\xi^{2}}\biggr)\xi^{4},\end{equation}
for the solution with the minus infront of the square root.

For comparison we present here the dispersion of plane wave in
unpolarized ultracold Fermi gas in introduced above dimensionless
variables
\begin{equation}\label{DF3D disp dep dimless unpol Fermi}\Omega^{2}=\frac{1}{3}(3\pi^{2})^{2/3}\biggl(1+32\gamma\biggr)\xi^{2}+\biggl(\frac{1}{4}+3\gamma\biggr)\xi^{4},\end{equation}
where following paper ~\cite{Andreev PRA08} we include
contribution of SRI between Fermi particles up to the third order
by the interaction radius.

We can see that solution (\ref{DF3D disp dep dimless small pol Fermi plus})
differs from (\ref{DF3D disp dep dimless unpol Fermi}) by one term, the
last one proportional to $d$. In the case of particles having no electric
dipole moment $d=0$ we have that the solution (\ref{DF3D general disp dep dimless})
with plus infront of the square root describes well-known matter wave in
ultracold Fermi gas, and formula (\ref{DF3D disp dep dimless small pol Fermi minus})
presents dispersion dependence of new wave appearing due to polarization dynamics.

Introducing the average distance between particles
$l=1/\sqrt[3]{n_{0}}$ we can rewrite definition of $\xi$ in the
following form $\xi=2\pi l/\lambda$, where $\lambda$ is the
wavelength $\lambda=2\pi/k$. As wavelength must be at least more
than two distances between particles we have that $\xi<1$ or, in
more realistic cases, we should consider $\xi\sim$0.1-0.001. Thus
we can admit that $\xi^{2}$ on several orders smaller than one. To
get $\gamma=1$ we need $\mid\Upsilon_{2}\mid>10^{-20}$ $\sqrt{\mu g}(\mu m)^{5.5}/c$.
For equilibrium particle concentration equal to $10^{14}$
cm$^{-3}$, particle mass $m=2\cdot 10^{-23}$g and electric dipole
moment of particle $p=1$ Debuy we have $d=0.8$. More interesting for
future experiment case when molecule has electric dipole moment
equal to 100 Debuy, in this case $d\simeq 10^{4}$. Formula (\ref{DF3D
general disp dep dimless}) presents two solutions, for the case
when particles have large electric dipole moment for solution with
the sign plus in front of square root we get
\begin{equation}\label{DF3D disp dep dimless appr with plus}\Omega=\sqrt{4\pi\sigma d}\xi,\end{equation}
or in the other terms
$\omega=\hbar  \sqrt[3]{n_{0}}\sqrt{4\pi\sigma d}k/m$, this
solution appears at neglecting all terms in comparison with two
terms proportional to $\sigma d$. Let's consider the second
solution presented by formula (\ref{DF3D general disp dep
dimless}) with sign minus in front of the square root. If we left
two terms only, which proportional to $\sigma d$ we get
$\Omega=0$. Thus, we have that two large terms reduce each other,
so to get the solution we should consider other terms. Making
expansion of the square root in linear on $1/d$ approximation we
get dispersion of the second mode
$$\Omega^{2}=\xi^{2}\Biggl(\frac{1}{3}(3\pi^{2})^{2/3}\biggl(1+32\gamma\biggr)$$
\begin{equation}\label{DF3D disp dep dimless appr with minus}+\biggl(\frac{1}{4}+3\gamma-\frac{2}{15}\frac{\kappa}{\sigma}(3\pi^{2})^{2/3}\biggr)\xi^{2}\Biggr).\end{equation}
From comparison of formulas (\ref{DF3D disp dep dimless
unpol Fermi}) and (\ref{DF3D
disp dep dimless appr with minus}) we can see that in the limit of a large equilibrium
polarization dispersion dependence of the "second" matter wave
similar to the dispersion of matter wave in unpolarized ultracold
Fermi gas (\ref{DF3D disp dep dimless unpol Fermi}), with only
difference - existence of the last term in formula (\ref{DF3D
disp dep dimless appr with minus}).

Following Ref. ~\cite{Andreev arxiv 12 transv dip BEC} we can admit that account
of contribution of the transverse electric field in dynamics of matter waves
in electrically polarized ultracold Fermi gas should lead to replacement
of $d$ by $d \cos^{2}\theta$. Thus, we can write that in general case we get
\begin{equation}\label{DF3D disp dep dimless appr with plus anisotropic} \Omega=\sqrt{4\pi\sigma d}\cdot |\cos\theta |\cdot\xi\end{equation}
instead of (\ref{DF3D disp dep dimless appr with plus}) for dispersion of
the collective excitations in electrically polarized ultracold Fermi gas
of molecules with large electric dipole moment. We can also admit that made
replacement has no influence on solution (\ref{DF3D disp dep dimless appr with minus}).

Here we have considered dispersion dependence appearing from dynamics of $\delta n$, $\delta v_{x}$, $\delta P_{x}$, $\delta R_{xx}$ and $\delta E_{x}$.
Quantities $\delta P_{y}$, $\delta R_{yx}$ and $\delta P_{z}$, $\delta R_{zx}$ give us two pair of independent sets of equations. Both of them give same dispersion relation:
$$\omega^{2}=\frac{\hbar^{2}k^{4}}{4m^{2}}+\kappa (3\pi^{2}n_{0})^{2/3}\frac{\hbar^{2}}{5m^{2}}k^{2}.$$

Taking into account $E_{0z}\neq 0$ (or $E_{0y}\neq 0$) gives no influence on evolution of $\delta n$, $\delta v_{x}$, $\delta P_{x}$, $\delta R_{xx}$ and $\delta E_{x}$. However it gives contribution in evolution $\delta P_{z}$, $\delta R_{zx}$. $\delta n$ gives influence on $\delta P_{z}$, $\delta R_{zx}$, but in this case we have not closed set of equations.

Now, we have described main properties of dispersion of collective excitation
and we can pass to conclusions.

\begin{figure}
\includegraphics[width=8cm,angle=0]{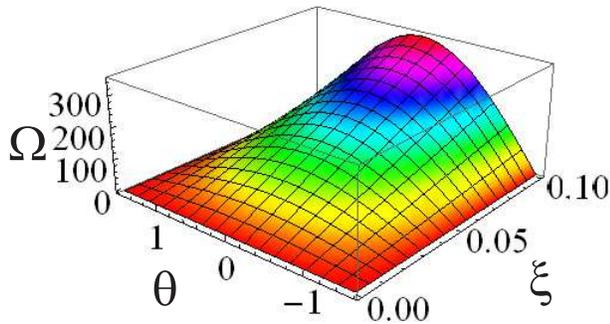}
\caption{\label{dip Fermi parallel} (Color online) The figure describes
the dispersion dependence of the collective excitation in the polarized
ultracold Fermi gas presented by formula (\ref{DF3D disp dep dimless appr with plus anisotropic})
at $d=100$ and $\sigma=1$.}
\end{figure}

\section{\label{sec:level1} IV. Conclusion}

We have developed the method of the many-particle QHD for the
system of the ultracold electrically polarized Fermi particles.
The set of the QHD equations consists of four material equations:
the continuity equation (the equation of balance of the particles
number), the Euler equation (the momentum balance equation), the
equation of polarization evolution, and the equation of
polarization current evolution, \emph{and} two equations of field,
which is the part of the set of the Maxwell's equation. Due to the
fact of using of the mentioned equations of field we can admit
that we have deal with the longitudinal waves, i.e. direction of
the electric field in the wave parallel to the direction of wave
propagation.

Using the QHD equations we have studied dispersion of the plane
collective excitation in the three dimensional ultracold
electrically polarized Fermi gas. We have found that there are two
matter waves, whose dispersion reveals no anisotropy. These two
waves appear in the electrically polarized Fermi gas instead of
the one matter wave existing in the unpolarized Fermi gas. Using
our previous results we generalized the obtained dispersion
dependencies to get anisotropic spectrum.

\section{\label{sec:level1} Acknowledgments}
The author thanks Professor L. S. Kuz'menkov for fruitful discussions.

%\nocite{*}

\end{document}